\documentstyle[psfig]{l-aa}
\begin{document}
\thesaurus{08.02.3, 08.03.3, 08.03.5, 10.15.2}
\title{Optical Identification of ROSAT sources in M~67:	 activity in an
old cluster
\thanks{Based on observations collected at ESO, La Silla}}
\author{L. Pasquini \inst{1}
\and T. Belloni\inst{2} }
\offprints{Luca Pasquini}
\institute{ European Southern Observatory, Karl Schwarzschild Str. 2,
85748 Garching bei M\"unchen, Germany
\and Astronomical Institute ``Anton Pannekoek'', University of
Amsterdam and Center for High-Energy Astrophysics, Kruislaan 403,
NL-1098 SJ Amsterdam, The Netherlands}
\date{Received; Accepted}
\maketitle
\begin{abstract}

We present optical identification and high-resolution spectroscopy
of ROSAT sources in the field of the old open cluster M~67.
For the first time it is possible to analyze coronal and
chromospheric activity of active stars in a solar-age cluster, and to
compare it  with  field stars.	ROSAT observed the high
X-ray luminosity tail of the cluster sources. In agreement with
what expected from studies of field stars,
most of the detected X-ray sources are binaries,
preferably with short periods and eccentric orbits.
In addition, several of the M~67 ROSAT sources have peculiar
locations in the cluster colour-magnitude diagram. This
is most likely due to rather complex evolutionary histories,
involving the presence of mass transfer	 or large mass losses.

The X-ray luminosity of the sources does not scale
with the stellar parameters in an obvious way.
In particular, no relationship is found
between coronal emission and stellar magnitude or binary period.
The Ca II K chromospheric flux from most of the counterparts is
in excess to that of single stars in the cluster by one order of magnitude.
The X-ray luminosity
of the sources in the old M~67 is one order of magnitude
lower than the most active active binaries in the field, but
comparable to that of the much younger binaries in the Hyades.


\end{abstract}

\section{Introduction}

The study of chromospheric and coronal activity has progressed impressively
in the last two decades, thanks to the launch of X-ray satellites.
Coronal sources have been detected in X rays all over the cool part of
the H-R diagram (Vaiana et al. 1980) and active binaries of the RS CVn
type  have soon been recognized as powerful X-ray emitters
(Walter et al. 1980).

With the advent of the ROSAT satellite and its all-sky survey, a complete
X-ray study of known active binaries  has been possible
(Dempsey et al. 1993a,b). A large number of previously unknown
coronal active stars is becoming available
(see e.g. Metanomski et al. 1998), with interesting
consequences on our understanding of the population of
young stars in the solar neighborhood (Guillout et al. 1998).

While the general framework of dependence of X-ray emission on
stellar rotational velocity (and then stellar age)
has been established long time ago
(Pallavicini et al. 1981), there still remain many uncertainties
about which stellar parameters affect the coronal emission in late-type stars.

We know, for instance, that old stars may preserve a high level of X-ray
emission if they are in binary systems, where a high rotational velocity
can be maintained through tidal interaction, but it is not clear which
role stellar mass, radius, orbital period
and eccentricity play in determining the level of activity.

One of the most relevant limitations is that,
since most studies are performed in field
samples, it is difficult to determine precisely the
characteristics of the studied objects. Stars with different
(but poorly determined) masses, ages and possibly evolutionary
histories are often compared.

The study of clusters, where ages, masses, and evolutionary status
of the counterparts of X-ray sources can be well established, is a necessary
step forward. In this framework, not only
ROSAT observations of young clusters have
provided new insight to the discussion on the age-activity relationship
(see e.g. Randich and Schmitt 1995), but for the first time
it has also been possible to detect coronal sources
in old clusters, i.e. with ages comparable with the Sun.

With the ROSAT observations of M~67 and other
intermediate age clusters, Belloni et al.(1993, 1996, 1997)
allowed to extend the study of the evolution of coronal activity to systems
with age up to $\sim$ 6 Gyrs. By examining these
coeval samples, these observations allow the study of
active coronae from a different, unique perspective:
it is possible to
investigate which stars (or stellar systems) show the highest
activity level in samples  so old that
the emission from single, solar-type stars is expected to be very low.

The ROSAT observations require a follow up
at different wavelengths, in order to :

\begin{itemize}

\item Confirm the identification of the optical counterparts.
In these clusters the stellar density is rather high,
and more than one source may be contained in the ROSAT error box.
As an example of the relevance of this issue,
the reader can compare the results of the
present work (summarized in Table 1) with those of Belloni, Verbunt 
\& Schmitt (1993),
based only on X-ray and general optical information, but not
on a detailed follow up.

\item Study in more detail the characteristics of the
sources, their binary nature, X-ray and chromospheric emission.

\item Compare the observed characteristics with those of field stars.

\end{itemize}

In this work, we present the identification and the study of
the ROSAT counterparts of the old open cluster M~67 (Belloni,
Verbunt \& Schmitt 1993).
A second, longer ROSAT pointing to M~67 has been performed (Belloni,
Verbunt \& Mathieu in preparation):
where possible, the results from this observation have
been included in the present paper.

M~67 is a very interesting target, not only because its age and metallicity are
similar to those of the Sun, but also because the cluster has been subject to
several photometric and spectroscopic studies which have led to
detailed membership determinations (Sanders 1977, Girard et al. 1989),
a large amount of data on binary stars (Latham et al. 1992),
photometric variables, W UMa candidates (Gilliland et al. 1991), and
blue stragglers (Mathys 1991).

This large body of  ancillary data makes M~67 one of
the best studied clusters;
one AM Her system and a hot white dwarf belonging to the cluster
have already been confirmed within this identification project (Pasquini
et al. 1994a).

\section{The observations}

The observations were carried out at ESO, La Silla, over the period 1992-1995,
using a variety of instruments and telescopes.
We stress that, being mainly
interested in the coronal counterparts of ROSAT detections,
we did not attempt a complete identification of all the sources, but
only of the possible coronal cluster members.
This implies that, when only optically faint candidates
are within the ROSAT error box, observations were not always pursued
further.

First, observations were carried out at the ESO 1.52m telescope equipped with
the B\&Ch spectrograph (Turatto 1997): with a resolution of
2 {\AA}/pixel and a 2048 pixel	CCD, the range 3700-7600
{\AA} was covered. All objects within the reach of the
telescope-spectrograph combination were observed, within $\sim$ 40$''$
from the nominal ROSAT source in order to be confident that no possible
counterpart would be missed.
A posteriori, we found this radius exceedingly large and we could confirm the
good accuracy of the ROSAT error box.

Bright stars within fields with no obvious counterpart have been
observed even if located at a comparatively large distance from the
X-ray position.

The low-resolution spectra were inspected to derive
(or check) spectral types or spectral anomalies, but mostly to find
signatures of high chromospheric emission, like filling-in or emission
of the Balmer lines. Most of the stellar candidates were
selected in this way. Additional observations at low resolution needed for the
identification of some of the fainter counterparts were performed using
the 3.6m telescope with the EFOSC spectrograph (Benetti et al. 1997;
see also Pasquini et al. 1994a).

Finally, the 3.6m telescope with the CASPEC spectrograph (Randich and
Pasquini 1996) and the NTT
with the EMMI spectrograph were used to obtain intermediate- and
high-resolution spectroscopy of the pre-selected candidates.
This last step is required to
allow a firm identification of the targets and to derive absolute
chromospheric fluxes at the stellar surface. The  observations were
centered in the Ca II H and K region, and in the H$\alpha$ region.
CASPEC spectra have a resolving power  R=18000.
The EMMI spectra were obtained	in
dichroic mode: blue and red spectra were recorded simultaneously.
Blue spectra were acquired with a holographic grating at R=6000
(Pasquini et al. 1994b), while red spectra were obtained with a resolution
R=20000 or R=3000, depending on	 the stellar apparent magnitude.
The spectra have been reduced using the MIDAS package (Banse et al. 1988).

When a candidate showed enhanced chromospheric activity and acceptable
positional coincidence, it was accepted
as counterpart. However, for some of the fields no star fulfilled
both conditions.
For these fields, we obtained low resolution spectroscopy of fainter
candidates, and when no acceptable alternatives were found, the original stars
were accepted as (possible) candidates.

\section{Optical Identification: results}

Belloni, Verbunt \& Schmitt (1993, hereafter BVS) list 22 sources detected
in the central 20 arcmin of a ROSAT PSPC field
centered on a position $\sim 12'$ east of the center of M~67.

The ROSAT sources with firm or possible optical
 counterparts  are summarized in Table 1:
the numbering scheme is the same as in BVS.
Source number 11 was recognized by BVS as a blend of
a hard and a soft source.
In the analysis of the new ROSAT observation of M~67 (Belloni, Verbunt
\& Mathieu, in preparation), the two sources could be
resolved and they are listed separately (sources 11a and 11b).
The optical positions are from Girard et al. (1989)
when available, while the coordinates for the remaining objects
are from the Digitized Sky Survey.
For these, distances are  accurate only to a few arcsecs.

For optical identification, we adopted the results obtained by similar
studies in the field (see e.g. Stocke et al. 1991).
QSO's and emission-line galaxies, when
close to the nominal error box were accepted as
counterparts; in particular, in the error boxes of the 3 extragalactic
sources identified with X-ray sources, no known M~67 member exists.
For the ten ROSAT fields
with no identification, EFOSC images were acquired,
but no stellar counterparts was found down to magnitude limits in
excess of those expected for values of L$_x$/L$_v$
typical for coronal sources.

In Figure 1, low resolution spectra of the extragalactic sources and
of some of the most interesting objects
in the fields with doubtful or no optical identification are shown.
Note that the shape of the continuum  may not represent
the true continuum of the objects, since in order
to gain spectra of several objects simultaneously,
in most cases the slit could not be aligned
with the parallactic angle.

\begin{figure}
\psfig{figure=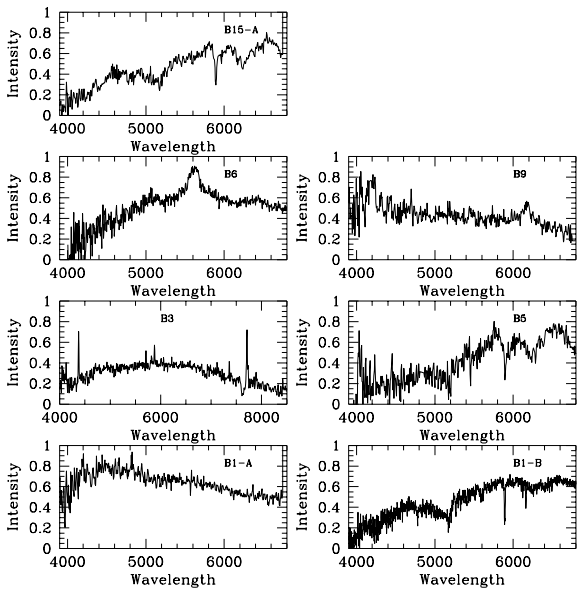,width=9.0cm,silent=}
\caption[1]{Low resolution spectra of the extragalactic 
optical counterparts and of the objects in the fields with doubtful or
no optical identification. Flux is in arbitrary units}
\end{figure}

Thanks to the boresight correction applied by BVS, the optical
and X-ray positions for most candidates agree extremely well, with
differences often smaller  than 10 arcsec, which is the typical nominal
ROSAT error box (BVS).
Therefore, we consider the 2 proposed
counterparts having larger distances from the X-ray source (B1, B15)
only as possible, but not likely candidates.
For B19, the uncertainty in the identification is given by the
fact that, although the X-ray and the optical position of the star
S364 match quite well, the high-resolution spectra obtained do not show
any sign of enhanced chromospheric activity (cfr. section 5.2).
This, according to the previously mentioned
criteria, does not make this star a firm counterpart.

\begin{table*}
\caption{Summary of the identification of ROSAT sources.
For each X-ray source we list: identifier, name of the
possible optical counterpart, distance between them, 90\% error
radius (from BVS), and a comment on the object.}
\parskip0.2cm
\begin{center}
\tabcolsep0.2cm
\begin{tabular}{r|r|l|l|l}\hline\hline
\multicolumn{1}{c}{B\#} &
\multicolumn{1}{c}{Name} &
\multicolumn{1}{c}{d ($''$)} &
\multicolumn{1}{c}{r ($''$)} &
\multicolumn{1}{c}{Comments}\\ \hline
1   & ---      & 38   & 17.8 & DISTANT \\
3   & ---      & 17   & 14.4 & Galaxy Z=0.174  \\
4   & S 1082   &  9   & 10.5 & Blue Straggler\\
6   & ---      &  4   &	 7.8 & QSO Z=1\\
7   & S 1077   &  4   &	 8.2 & \\
8   & S 1063   &  2   &	 6.6 & \\
9   & ---      & 22   & 18.3 & QSO Z=1.2 \\
10  & S 1040   &  7   & 12.5 & \\
11a & S 1019   &  7   & 14.4 & \\
11b & G 152    &      & 14.4 & WD (Pasquini et al. 1994a) \\
13  & S 999    &  3   & 14.2 & \\
14  & S 759    &  6   & 13.0 & \\
15  & ---      & 31   &	 7.8 & DISTANT \\
16  & G 186    &      & 15.5 & AM Her (Pasquini et al. 1994a) \\
17  & S 972    & 12   & 20.8 & \\
19  & S 364    & 16   & 16.0 &	POSSIBLE \\
\hline
\end{tabular}
\end{center}
\end{table*}

Out of 23 sources,
one has been recognized as spurious, and 13 have been
firmly identified. Of these, 9 are cluster members,
3 are extragalactic objects and one is a non-member star.
Three additional cases remain unclear, in the sense that they do
not meet all the requirements for a firm identification.

As explained in detail below, of the identified sources, 
seven are definitely coronal emitters and belong to the cluster.
Relevant for the aim of this work is that
we can be confident that none of the unidentified sources
are coronal emitters belonging to the cluster.
This confidence is supported by three arguments:

\begin{itemize}

\item All unidentified sources
have faint optical counterparts. This would imply an
exceptionally high L$_x$/L$_v$ ratio if they were coronal sources. 
Only in the case of B12
there are possible coronal candidates
the error box, but their cluster membership is uncertain.

\item Most of the ROSAT detections in the central part of the cluster have
been identified; the unidentified sources are located in the external
parts of the cluster, where the membership probability is lower.

\item As pointed out by BVS,
on the basis of the LogN-LogS function (Hasinger et al. 1998), 
$\sim$ 12-16 sources out of the 23 detected sources are expected to be
background objects.

\end{itemize}

However, we cannot exclude that
some non coronal system belonging to M~67
are present among the unidentified ROSAT sources.
Systems with a high L$_x$/L$_v$ ratio, like white dwarfs (WD),
cataclysmic variables (CV), or
low mass X-ray binaries (in quiescence) could be present,
but not identified in the present program.
WD's and polar CVs are very unlikely cases, since
they would show very soft hardness ratios (Fleming et al. 1996),
which are not observed (BSV), but we cannot exclude the presence
of a weakly-magnetized CV or a LMXB in quiescence.

Below, we discuss the single sources in detail:

B1: The proposed possible counterpart (star B1-B in Figure 1)
is not in the list by Sanders (1977). Its coordinates are:
$\alpha$(2000)= 08 49 53  $\delta$(2000)= 12 02 06.
At 2 {\AA} resolution the star shows H$\alpha$ filling-in.
No other star in the field shows signatures of activity.
We derive a spectral type K5: with a
magnitude of V$\sim$16.4, the star could be a member of the cluster.
However, it is distant from the X-ray position.
A stellar object of magnitude V$\sim$20.4 and position
$\alpha$(2000)= 08 49 54
$\delta$(2000)= 12 01 48, lies only 16 arcsec away from the X-ray position
(B1-A in Figure 1). Its spectrum resembles that of an F dwarf
(although with unusually faint Balmer lines).
With this combination of magnitude and spectral type, the star does not
belong to the cluster, and the exceedingly
high L$_x$/L$_v$ would exclude it as a coronal counterpart.

B2: Two stars are found at ($\alpha$(2000)=08 50 14.8; $\delta$(2000)
= 12 00 37) and ($\alpha$(2000)=08 50 14.6; $\delta$(2000)= 12 01 45),
at a distance of 28 and 42 arcsec from the X-ray position
respectively. Their spectra are those of a M1 and a K4 dwarf with no special
features. An additional object lies very close to the
X-ray position, possibly with and emission line at $\sim$6500 {\AA},
but the signal to noise ratio is too low to derive
any firm conclusion.

B3: A galaxy with coordinates $\alpha$(2000)= 08 49 22.6
$\delta$(2000)= 11 54 57.6.
The  two strongest emission lines are consistent with being [OII]3727
and H$\alpha$ at redshift 0.174 (see Figure 1).

B4: This field has not been extensively studied. The agreement of S1082
with the X-ray coordinates is very good and the brightness of the star makes
the study of nearby faint objects difficult.
Low resolution spectra for stars
S1072, S1075 and S1079	were obtained. They do not show signatures of
enhanced activity.
The high resolution spectrum of S1082 shows a composite
H$\alpha$, with a broad and a narrow component (cfr. Figure 2). Such a
spectrum was also observed by Mathys (1991).

B5: A faint object is in good agreement with the X-ray
position ($\alpha$(2000)=08 50 08.8 $\delta$(2000)=11 53 27, d=15 arcsec). 
However, from visual magnitude and spectral type of the star
(M1 V, see Figure 1), Balmer lines in emission are expected
for it to be the optical counterpart. The nearest S star (S490)
shows no special features. The counterpart is therefore not confirmed.
Among all the BSV sources, this is the only one
which has not been confirmed by the new M~67 pointings
(Belloni, Verbunt \& Mathieu, in preparation) and it should probably be 
considered spurious.

B6: A QSO with coordinates $\alpha$(2000)=08 49 54.2
$\delta$(2000)= 11 53 17.
The redshift has been computed identifying the strong emission line as Mg II.
(see Figure 1).

B7: S1077 shows indications of activity. At low resolution the nearby 
S2224 does not show any peculiar signature.

B8: S1063 not only coincides very well with position, but also shows a
very pronounced Ca II H \& K emission. This star is located below the subgiant
branch (cfr. section 5.3).

B9: The bright star in the field (S258) was observed at high resolution: it
does not show signs of activity, and it lies at a large distance from the X-ray
position. A QSO at Z=1.2 is found at $\alpha$(2000)=08 49 36,
$\delta$(2000)=+11 50 56.
The two emission lines (see Figure 1) correspond to CIII and MgII.

B10: S1040 shows Ca II H\&K emission. The fainter star
S2216 does not show particular features at low resolution.

B11a: The most difficult field: a crowded region, with several known
binaries. The best candidate is	 S1019, with spectacular Ca II H\&K emission,
and H$\alpha$ filling. Other brighter candidates have been observed also at
high resolution, but do not show evidence for enhanced activity
(S1010, S1045, S1024) (However, note that, even though it does not show Ca II 
emission, the Ca II flux of S1024 may still be quite high because to its 
rather hot temperature, cfr. Table 3).

B12: Two stars lie within 25 arcsec from the ROSAT position:
S613 (V=15.45) and S614 (V=15.50), but
neither shows peculiar characteristics at low resolution. Their membership
probability is given as low/intermediate by Girard et al. (1989)
(47 and 60 $\%$ respectively) and by Sanders (47 and 22 $\%$ respectively).
We do not consider them as good candidates. High resolution spectroscopy
is needed to reach a firm conclusion.

B13: S999 shows Ca II core emission at high resolution. At low resolution,
neither of the other possible candidates (S995, S997, S998) show special 
features.

B14: The Ca II spectrum of S759 shows enhanced emission, but the S/N ratio
of the spectrum is not very high. The nearby S756, S757 do not show special
features at low resolution. S759 is given zero probability of membership
by Sanders (1977).

B15: A late K star. At intermediate resolution, we could only obtain a
low S/N spectrum: the star might show Ca II H\&K 
emission, but the distance
from the X-ray source makes the identification uncertain. A fainter star
(V$\sim$18.5) at $\alpha$(2000)= 08 50 55.9 $\delta$(2000)= 11 45 54
(d=17.5 arcsec) shows an M1 dwarf spectrum (see Figure 1), but the absence of
Balmer-line emission makes it an unlikely counterpart.

B16: AM Her (Gilliland et al. 1991, Pasquini et al. 1994a)

B17: Two stars (S972 and S973) are close to the X-ray source: intermediate and
high resolution spectra show strong Ca II H\&K in S972 but not in S973,
which is a brighter (V=13.49) binary (Orbital Period=40.4 days, Latham et
al. 1992). S972 is listed by Sanders (1977) with a moderate (42\%) probability
of membership.

B18: No spectroscopic observations.

B19: The bright giant S364 ($\alpha$(2000)= 08 49 57 $\delta$(2000)=11 41 36)
is within the ROSAT error box. Its membership probability is listed as 82\%
by Sanders (1977). Low resolution spectra were obtained for
two other possible counterparts (no S numbers) distant 47 and 24 arcsec
respectively. Neither shows special features. The fact that S364 does not show
any enhanced activity either in the Ca II K or in the H$\alpha$ lines
(see Figure 2 and Section 5.2) makes its identification with the ROSAT
source uncertain.

B20-22: No spectroscopic observations. No stars bright enough to be
obvious coronal counterparts are visible in the fields.

\section{Intermediate- and high-resolution Spectra}

The Ca II and H$\alpha$ high-resolution spectra are shown in Figure 2.
It is relevant to note that we observed at high resolution not only the 
proposed optical candidates, but also several stars
with similar characteristics.
In Table 2, the optical characteristics of the coronal optical
counterparts are
given, together with the X-ray luminosity. The
X-ray luminosities differ by a factor $\sim$ 2 from those given by BVS,
because in this early work a too low conversion factor was
used. Here we adopt a the same model of Belloni, Verbunt \& Mathieu 
(in preparation)
with N$_H$=1.7$\times$10$^{20}$cm$^{-2}$. This corresponds to a
 conversion factor between counts/sec and 0.1--2.4 keV flux in 
erg cm$^{-2}$sec$^{-1}$ of 1.8$\times$10$^{-11}$.

\begin{figure}
\psfig{figure=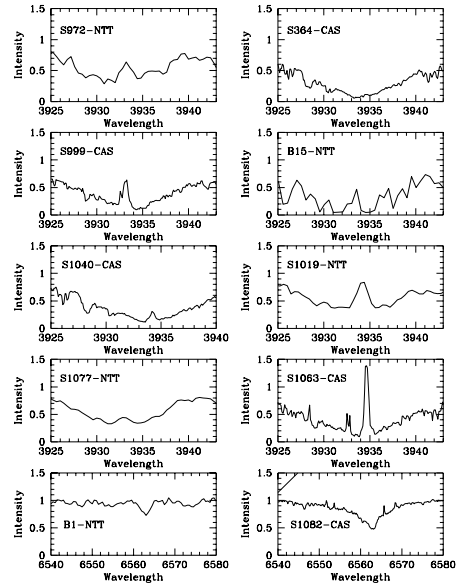,width=9.0cm,silent=}
\caption[2]{ High- and intermediate-resolution Ca II and H$\alpha$
spectra of the proposed candidates}
\end{figure}

The properties of the `comparison' stars observed at high
resolution are summarized in Table 3. Some of these stars
turned out to be likely detected in the new M~67 observations of Belloni,
Verbunt \& Mathieu, in which case the X-ray luminosities and fluxes are
given for sake of completeness.
However, since these detections refer to a pointing having different
sensitivity and the
analysis of the optical counterparts has not yet been performed as in
the present analysis, they are treated separately from the stars of
Table 2, which form the bulk of the present work.

We used the high-resolution spectra to compute
Ca II K line fluxes using the procedure of Pasquini et
al. (1988), based on the (V-R) colour index.
For stars having (V-R) colours measured by Gilliland
et al. (1991) in the Cousin system, the Johnson (V-R)$_o$
colour has been derived by using the transformation
formulas given by Taylor (1986). For the other stars, the (V-R)$_o$ colour
has been computed using (B-V)-(V-R) standard transformations (Johnson 1966).

For a few stars, both high (R=20000) and
intermediate (R=6000) resolution spectra were acquired.
Chromospheric fluxes were  computed by using both
high and intermediate resolution spectra and the results were compared
to evaluate the effects of the different resolution
on the flux estimates. The A$_k$ indices
(the ratio between the Ca II core emission and the pseudocontinuum
at 3950 {\AA}, see Pasquini et al. 1988) derived from low resolution spectra are
systematically larger by a factor 2 than those measured from the
high resolution ones. Such an effect is expected, as discussed in
Pasquini et al. (1989). The A$_k$ values given in Tables 2 and 3 have been
corrected for this difference. The uncertainties in the chromospheric fluxes
are of the order of 50$\%$. For a few stars, marked with `:' in Tables 2 and 3,
the uncertainties may be even higher, due either to the low signal-to-noise
ratio of the data, or to problems in defining accurately the K1 minima.

In Figure 2, high and intermediate resolution Ca II K line spectra
are shown.
The spectra are normalized to the 3950 {\AA} pseudocontinuum.

In Tables 2 and 3, the derived chromospheric fluxes are also given. No
attempt was done to separate the different components of a binary system, 
since they are never clearly separated in our spectra.

\begin{table*}
\caption{Summary of the ROSAT sources with identified optical counterparts.
Column 1): ROSAT ID, Column 2): Sanders (1977)
number, Column 3): V magnitude,
Column 4): B-V colour, Column 5): V-I colour, Column 6):
V-R colour, Column
7): Orbital Period (Latham et al. 1992), Column 8): Photometric period, Column
9): Log of X-ray Luminosity, Column 10): Chromospheric index in the Ca II K 
line, Column 11): Chromospheric flux in the Ca II K line in units of 10$^5$ 
erg cm$^{-2}$ sec$^{-1}$. Column 12): Angular diameter in milliarcsec 
according to the relationship by Barnes and Evans (1976), Column 13): Stellar 
radii in units of 10$^{10}$ cm.
$^a$ The star is on the binary sequence of Montgomery et al. (1993),
$^b$ Photometric period from Goranskij et al. (1992), $^c$ Photometric period
from  Gilliland et al. (1991). For the derivation of the
(V-R) colours and the chromospheric fluxes see text. }
\parskip0.2cm
\begin{center}
\tabcolsep0.2cm
\begin{tabular}{r|r|r|r|r|r|r|r|r|r|r|r|r}\hline
\multicolumn{1}{c}{Source} &
\multicolumn{1}{c}{Name}   &
\multicolumn{1}{c}{V}	   &
\multicolumn{1}{c}{B-V}	   &
\multicolumn{1}{c}{V-I}	   &
\multicolumn{1}{c}{V-R}	   &
\multicolumn{1}{c}{O.P.}   &
\multicolumn{1}{c}{Ph.P.}  &
\multicolumn{1}{c}{Log(L$_x$)}&
\multicolumn{1}{c}{Ak}	   &
\multicolumn{1}{c}{F'k}	   &
\multicolumn{1}{c}{Log($\phi$)} &
\multicolumn{1}{c}{R}  \\ \hline\hline
1  & No S\#&	    &	    &	   &	 &	&	   &30.61&/  &/	  & /	  & \\
4  & S1082  & 11.25 & 0.415 &0.529 &0.35*&	& 1.068$^b$&30.98&0.2& 40:&-1.4283&21.9\\
7  & S1077  & 12.47 & 0.637 &0.843 &0.50*&	&	   &30.85&0.5& 50 &-1.5436&16.8 \\
8  & S1063  & 13.52 & 1.07  &1.199 &0.85*& 18.24&	   &31.16&0.95&10 &-1.4533&20.7 \\
10 & S1040  & 11.55 & 0.88  &0.91  &0.67 & 42.83& 7.97$^c$ &30.79&0.27&7  &-1.21374&35.9 \\
11a& S1019  & 14.32 & 0.83  &0.997 &0.67 &	& 9.75$^c$ &30.72&0.75&26 &-1.76774&10.0 \\
13 & S999   & 12.60 & 0.78  &0.908 &0.62 & 10.06& 9.79$^c$ &30.68&0.64&30 &-1.46664&20.0 \\
14 & S759   & 16.17 & 0.747 &0.828 &0.56*&	&	   & ??	 &0.23:&14:& N.M. &  \\
15 &No S\#&	    &	    &	   &	 &	&	   &30.83&0.27:&  &/ & \\
17 & S972   & 15.37 & 0.89  &1.09  &0.67*& $^a$ &	   &30.54&0.5 &18 &-1.97774&6.2 \\
19 & S364   & 9.93  & 1.36  &1.30  &0.94 &	&	   &30.50&0.15&0.5&-0.65808&129 \\
\hline
\end{tabular}
\end{center}
\end{table*}

\begin{table*}
\caption{Additional M~67 stars observed at intermediate/high resolution for
comparison purposes with stars of Table 2.
Columns are the same as in Table 2.
 $^a$: combined colour and magnitude
from Latham et al (1992), $^b$ Photometric period from Gilliland et al. (1991)}
\parskip0.2cm
\begin{center}
\tabcolsep0.2cm
\begin{tabular}{r|r|r|r|r|r|r|r|r|r|r|r}\hline
\multicolumn{1}{c}{Name} &
\multicolumn{1}{c}{V} &
\multicolumn{1}{c}{B-V}&
\multicolumn{1}{c}{V-I}&
\multicolumn{1}{c}{V-R}&
\multicolumn{1}{c}{O.P.}&
\multicolumn{1}{c}{Ph.P.}&
\multicolumn{1}{c}{Log(L$_x$)}&
\multicolumn{1}{c}{Ak} &
\multicolumn{1}{c}{F'k} &
\multicolumn{1}{c}{Log$\phi$} &
\multicolumn{1}{c}{R} \\ \hline\hline
S1010	&10.48	&1.11  &1.08  &.78   &	     &	       & /     &0.06 & 0.4:&-0.90536& 73 \\
S1024	&12.718 &0.553 &0.685 &.43   & 7.16  &	       &30.32: &0.12 & 10: &-1.65326& 13 \\
S1045	&12.54	&0.591 &0.703 &.46   & 7.645 &	       &29.82 &0.07 & 3	  &-1.59192& 15 \\
S1113	&13.766 &1.013 & //   &0.78* &	     &	       &30.79  &1.58 & 28  &-1.56256& 16.1\\
S1242	&12.72	&0.683 &0.807 &0.53* & 31.78 & 4.88$^b$&29.96  &0.20 & 14  &-1.56786& 15.9\\
S1264$^a$& 11.7 & 0.92 & //   &0.69* & 353.9 &	       & /     &0.09 & 2.2 &-1.22658& 34.8 \\
S1272	&12.514 &0.598 &0.697 &0.47* & 11.02 &	       & /     &0.11 & 10  &-1.57814& 15.5 \\
\hline
\end{tabular}
\end{center}
\end{table*}

\section{Discussion}

The aim of this work is to discuss the nature of the coronal sources
belonging to M~67 detected in the ROSAT observation of BVS. For this purpose
we concentrate mostly on the firmly identified sources.

\subsection{Which parameters determine X-ray emission in a 4Gyr Old Cluster?}

It is important to remember that the observations by BVS could only detect the
high X-ray luminosity tail of the cluster.
The longer, more recent ROSAT observations reveal more
sources,  confirming that cluster X-ray emitters
might exist at lower levels than the one analyzed here (Belloni, Verbunt \&
Mathieu, in preparation).

Of the 8 known members of Table 2, six are either known binaries or show
clear signs of duplicity, like photometric modulation or displaced
position in the colour-magnitude diagram of the cluster. Of the remaining two,
S1077 is reported by BVS as a multiple system
with short period, while S364 is not included in the current list of binaries.
Jones and Smith (1984) found anomalous DDO colours for this star,
but they concluded
that this was probably due to due measurement uncertainties.
Once again, we stress that this identification
is only considered as `possible'.

{\it The first conclusion is that
in M~67 the strongest coronal sources
are binaries}.	This was  expected from the results on field stars
and from the age-coronal activity relationships: single, late type
 stars with  ages comparable to the Sun are not expected to emit
X-ray in excess to $\sim$10$^{29}$ erg/s in luminosity.

In Figure 3 the colour-magnitude diagram of the cluster is shown, with the
X-ray identifications marked.
It is striking to notice that a large variety of cases exists:
one Blue Straggler (S1082), one peculiar object (S1063), evolved
`normal' binaries (S1077, S999),
main sequence binaries (S1019, S972), one Red Straggler
(S1040), and possibly a red giant (S364).
Among field stars such a comparison can hardly be done, due to the
uncertainty in the fundamental stellar parameters.

\begin{figure}
\psfig{figure=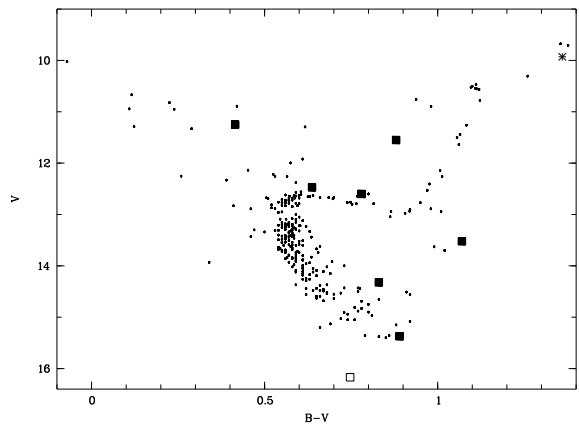,width=6.0cm,silent=}
\caption[3]{ Colour-magnitude diagram of M~67. Filled squares show the
positions of the firm X-ray counterparts. S759 (non member) and
S364 (possible counterpart) are indicated respectively as an
open square and a star. }
\end{figure}

{\it The second conclusion is therefore that among the M~67 sources there
exists a large
variety in evolutionary status and composition of binary systems.
In particular, several of the strong X-ray emitters are found among
objects having peculiar location in the colour magnitude diagram.}

This variety makes it difficult to understand which
parameters determine the X-ray emission in these binaries. From
the study of field active binaries, it emerged
that the main parameter determining X-ray luminosity is stellar radius.
Therefore, the strongest X-ray emitters
would be expected to be among the most luminous systems (Dempsey et al. 1993b).
On the contrary, the most luminous X-ray sources
in M~67 span over a range in visible luminosity of at least 4
magnitudes (cfr. Table 1).
Although  S972, which has the lowest X-ray luminosity, is also
the faintest established member of the cluster in the optical,
other stars (like S1019 or S1063) show the
highest X-ray emission, while being relatively faint in the optical.

{\it The third conclusion is therefore that,
in this coeval sample, the highest
X-ray luminosity is not given by the most luminous stars.}

Another interesting point is the dependence of X-ray emission on the
orbital (or rotational) period, since
X-ray luminosity scales with stellar rotational velocity
(Pallavicini et al. 1981), which in turn is related to stellar
radius and orbital period.
In most cases (and for main sequence stars), it is expected that
short period binaries
are synchronized, i.e. they have equal orbital and rotational periods.
It is also expected that synchronization happens before orbit circularization
(Zahn 1977). All 4 stars for which the rotational modulation is known,
as inferred from the photometric variability,
have rotational periods
of less than 10 days. For S1063, only the orbital period is known, but this
star may have a special history (see Section 5.3).
Pending more data on the rotational period of the other objects,
we can say that

{\it Not only the X-ray sources in M~67 are binaries, but also most of them
have short rotational periods.}

On the other hand, do all  M~67 short period systems show strong X-ray
(or chromospheric) activity ?
In M~67, 11 binaries have a measured orbital period shorter than 16 days
(Latham et al. 1992),
a canonical value for the definition of RS CVn systems,
but most of them were not detected in the ROSAT	observations.
Three of them (S1272, S1284, S1224) where out or at the very border of
the ROSAT field; one (S999) was detected, but the remaining
(S1045, S1234, S1024, S986, S1009, S1070 S1014, S810)
were contained in the ROSAT field but not detected.
These systems are likely active,
but with X-ray luminosities L$_x \le$ 30.3 (corresponding to the sensitivity
of BSV),  as it is suggested by
the likely detection of some of these stars in the deeper pointing
(S1045, S1024, S1234 and S1070, Belloni, Verbunt \& Mathieu, in
preparation).
Since some of the non-detected binaries, like S986, S1009 and S810,
have periods of 10 days or shorter,
the data collected up to now show that

{\it Among the binaries of M~67, a short orbital period does not
necessarily imply a high X-ray  luminosity}.


Since we are dealing mostly with evolved stars, for which short periods
can coexist with eccentrical orbits,
eccentricity could be a relevant parameter,
because the non-synchronization would
allow the possibility of having a rotational period shorter than the
orbital period. Of the detected binaries,
S1063 and S999 have highly eccentric orbits, but S1040 has an eccentricity
comparable with 0. This is surprising, because if the
photometric variability detected by Gilliland et al. (1991) really represents
the rotational period, than this star would be circularized but not 
synchronized. On the other hand, S1040 had probably a very complex
history, having had mass transfer in the past (Landsman  et al. 1997).
Considering this star as a special case	we, could argue
that non-circular binaries are favored among the strongest X-ray emitters.
However, in the list of Latham et al. (1992),
there are five short-period binaries with
eccentricities significantly different from 0 (S1284, S1272,
S1234, S1224, S1014). Two of them (S1014 and S1234) were in the ROSAT
field of view
but were not detected. Therefore, it seems that also a high eccentricity
does not represent {\it per se} a condition for strong X-ray emission.

\subsection{Chromospheric Activity}

The chromospheric flux of the Sun in the Ca II K line
is of 3-5$\times$10$^5$ erg cm$^{-2}$sec$^{-1}$ (Pasquini et al. 1988).
For the main sequence G stars in M~67,
being metallicity and age comparable with those of the Sun,
values comparable with this are expected.

Recently, Dupree et al. (in preparation) studied the chromospheric
emission in the Ca II lines for a sample of M~67 giants, using
calibrations and spectra of similar quality to the ones presented here.
In Figure 4 the Ca II chromospheric fluxes are plotted as a
function of the (V-R) colour, both for the stars in Dupree et al. 
(open squares)
and for the X-ray counterparts of Table 2 (filled squares).
The ROSAT stars have a level of chromospheric activity one
order of magnitude higher than the optically-selected giants in Dupree et al.
The only exception is  S364, which fits extremely well
within normal giants of similar spectral type.
For this reason, despite the reasonable
agreement with the X-ray position, this star is considered as a
doubtful counterpart.

\begin{figure}
\psfig{figure=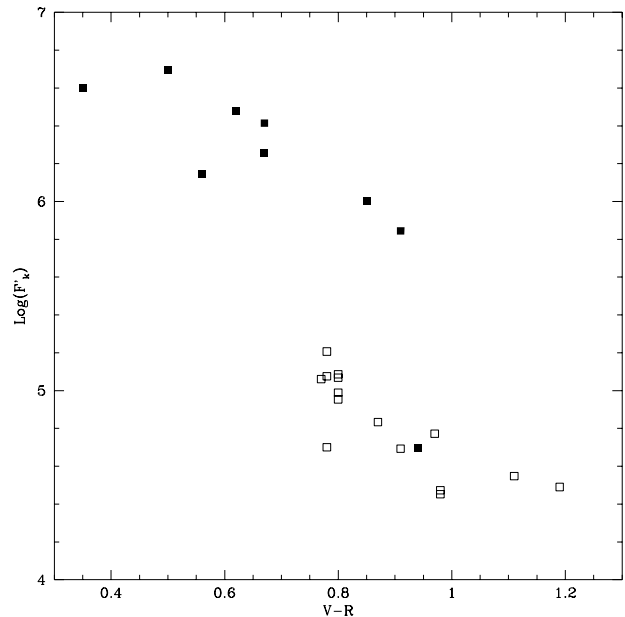,width=9.0cm,silent=}
\caption[4]{ Ca II chromospheric fluxes vs. (V-R) colour
for M67 single stars (Dupree
et al., in preparation: open symbols) and ROSAT stars (Table 2 of 
this work: filled symbols)}
\end{figure}

In order to perform an unbiased analysis of the
relationship between chromospheric and coronal activity,
we have to use indicators which express similar quantities,
namely fluxes at the stellar surface.
We computed the radii of the stars by using
the same Barnes-Evans (1976)
relationship used to calibrate the Ca II data:
Log$\phi$ = 0.4874 - 0.2V$_o$ + 0.858(V-R).
Although we know that the resulting radii are probably
incorrect (for instance, because the stars are implicitly assumed to
be single), the fact that the same
relationship is used for computing the Ca II and X-ray fluxes
minimizes the presence of possible systematic effects in the comparison.
We assumed an absorption A(V)=0.17  and a distance of
785 parsecs (Janes 1984).
The resulting diameters and radii are given in Tables 2 and 3
(in units of milliarcseconds and 10$^{10}$ cm respectively).
In Figure 5a, Log(F$_x$) is given as a function of
Log F$_k$. The relationship is rather scattered and
mostly the presence of the (doubtful) S364 hints to  the
presence of a trend.

Since the sample contains systems
whose evolutionary status is very different from each other, 
to further investigate this point 
we plot in Figure 5b the ratio between the X-ray and
chromospheric fluxes versus apparent magnitude.
Fainter (i.e. higher gravity) stars have much higher coronal to
chromospheric flux ratios than more luminous
(i.e. lower gravity) stars.

Although the number of objects is rather low, it appears that
dwarfs and giants follow different trends, with
dwarfs having higher F$_x$ for a given F'$_k$.
This fact has two possible explanations:

\begin{itemize}

\item For comparable chromospheric fluxes, higher-gravity stars are more
efficient in heating their coronae than lower-gravity stars.
This could indicate
the presence of different coronal structures between
dwarfs and evolved stars.

\item The assumption that the Ca II K fluxes are representative of the whole
chromospheric losses may not apply  when comparing stars of different
luminosity. Giants could for instance have a different balance
in the different chromospheric lines than dwarfs.

\end{itemize}

\begin{figure}
\psfig{figure=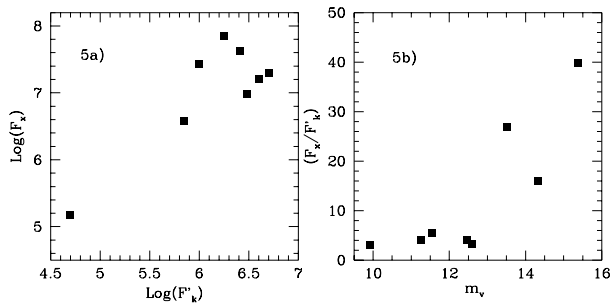,width=9.0cm,silent=}
\caption[5]{ 5a): Comparison between Ca II and X-ray fluxes
for the X-ray counterparts. 5b) Coronal to
Ca II fluxes as a function of stellar magnitude.}
\end{figure}

\subsection{S1063 and S1113}

Two objects deserve a special attention,
due to their peculiar position in the colour-magnitude
diagram: S1063 and S1113.

S1113 was was outside the field of the observation by BSV.
Because of the strong chromospheric activity observed, we
would expect it to show rather strong X-ray emission, 
which is indeed detected in the new
ROSAT pointing (cfr. Table 3).

These two stars are located below the
giants branch: they are as red as single subgiants, but almost
one magnitude fainter.
S1063 is a known eccentric binary with a period of 18.3 days, while
S1113 is a short period (2.82 d) circular binary (Latham et al. 
1992, Mathieu et al. in preparation).
The two stars are classified as  members in the
proper motion studies of Sanders (1977) and Girard et al. (1989),
with a probability higher than 90$\%$.
Their peculiar position in the colour-magnitude 
diagram and their high level of coronal and chromospheric 
activity make these two object very interesting.
It is not possible to simply combine two M~67 stars
and obtain the magnitudes and colours of S1063 and S1113.
Some mass exchange, or large mass losses
in the past history of the systems, possibly still going on, 
look unavoidable.
The high-resolution Ca II spectrum of S1113 shows neither direct
evidence of duplicity, nor strong asymmetries in the Ca II
core typical of strong mass losses, but not much can be derived 
with only one optical spectrum for this star.

We stress that the conclusion that these two stars have suffered a
special evolutionary history is made possible only by the
fact that we they are members of a cluster (and therefore we can firmly 
position them in the colour magnitude
diagram) and by the detailed optical follow-up. 
How many such systems exist among field binaries?
Note that, in absence of detailed studies
(e.g. accurate determinations of mass, metallicity and gravity),
similar systems in the field cannot be distinguished by otherwise
'normal' RS CVn binaries.
A more detailed study in other clusters and possibly
the analysis of Hipparcos parallaxes of active binaries will help in
understanding how common these systems are.
Investigation of their binarity and orbital synchronization
will also be crucial to model their possible evolution with time
(see i.e. the discussion in Stepien 1995).

\subsection{Comparison with ROSAT observations of field RS CVn}

One of the aims of this study is the possibility
of comparing for the first time
active stars in an old cluster with the field population of RS CVn.
Dempsey et al. (1993a,b) studied the ROSAT detections of known
RS CVn.
Their sample, taken from the catalogue of
Strassmeier et al. (1988), contains binaries of a large variety of ages,
masses and  periods.
We selected all the stars from Dempsey et al. (1993a)
with known distance. We computed X-ray luminosities
by converting the PSPC count rates (see also Dempsey et al. 1994)
using the same conversion factor used for the M~67 stars.
The distribution in luminosity is given
in Figure 6 for M~67 and field RS CVn separately. Although the M~67
sources do overlap well with the main body of field RS CVn's,
their emission is not as high as the most active RS CVn systems.

\begin{figure}
\psfig{figure=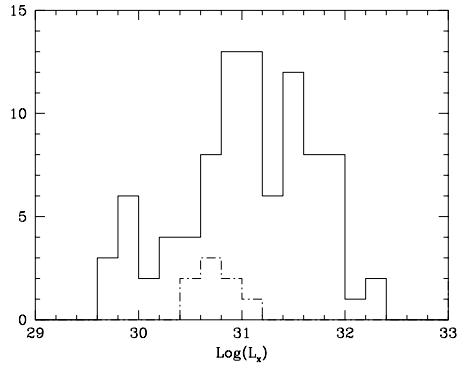,width=9.0cm,silent=}
\caption[6]{ ROSAT luminosity for the M~67 sources (broken
line) and for RS CVn binaries with known distances (continuous line)}
\end{figure}

{\it Figure 6 shows  that the high X-ray luminosity tail of binaries
in the 4 Gyr old cluster M~67 is about 10 times lower than the high
luminosity tail of field RS CVn's.}

The reasons for this difference are at the moment not very clear:
it could be an effect due to the old age of the cluster and/or to the
evolutionary status of the sources, or to a statistical effect
caused by the fact that we can only sample a few hundred
stars within the cluster. Identifications in clusters of different ages, 
as well as a detailed
analysis of the strongest sources in the field, will help in
understanding this open question. 
It is interesting to note 
that in the analysis of the much younger Hyades, Stern et al. (1995) 
found that binaries are the strongest X-ray emitters in this
cluster.
Similarities with M67 exist also in that the 
strongest X-ray source in the Hyades (V471 Tau) is a peculiar
system. 
Finally, Hyades binaries have X-ray luminosities similar or lower than
those observed among the  M67 sources. 

This could indicate that as far as the high-luminosity tail 
of the X-ray luminosity function
is concerned, age is not the primary parameter to 
determine the X-ray emission in binaries, 
as pointed out also by Ottmann et al. (1997)
in their analysis of Pop II binaries. 
 
We would therefore argue 
that the extremely active i.e. (L$_x$ $\sim$ 10$^{32}$ erg
cm$^{-2}$ sec $^{-1}$) RS CVn systems seem to be very rare and possibly
limited to quite exceptional cases. Metanomski et al. (1998), in their study of
hundred stars identified over a large area of the ROSAT All-Sky Survey,
find results similar to ours:
no such a high luminosity coronal source is indeed present in their sample.

\begin{acknowledgements}

We thank the referee, J.H.M.M. Schmitt, for many useful suggestions and 
comments.

\end{acknowledgements}

\end{document}